\documentclass{aa}
\usepackage{lineno}
\usepackage{graphicx}   % Including figure files
\usepackage{epsfig}
\usepackage{amsmath}    % Advanced maths commands
\usepackage{amssymb}    % Extra maths symbols
\usepackage{verbatim}
\usepackage{color}
\usepackage{txfonts}
\usepackage[version=4]{mhchem}
\usepackage{comment}
\usepackage{hyperref}
\hypersetup{
    colorlinks=true,
    linkcolor=blue,
    filecolor=red,
    citecolor=blue,
    urlcolor=blue,
}

\graphicspath{{./}{figures/}}

\begin{document}

   \title{Ammonia, carbon dioxide and the non-detection of the 2152 cm$^{-1}$ CO band}

   \author{Jiao He\inst{1,4}
         \and
         Giulia Perotti\inst{1}
          \and
          Shahnewaz M. Emtiaz\inst{2}
          \and
          Francis E. Toriello\inst{2}
          \and
          Adwin Boogert\inst{3}
          \and
          Thomas Henning\inst{1}
          \and
          Gianfranco Vidali\inst{2}
          }

   \institute{
    Max Planck Institute for Astronomy, K{\"o}nigstuhl 17, D-69117 Heidelberg, Germany
\and Physics Department, Syracuse University, Syracuse, NY 13244, USA
\and Institute for Astronomy, University of Hawai'i at Manoa, 2680 Woodlawn Drive, Honolulu, HI 96822–1839, USA
\and \email{he@mpia.de}
}

\authorrunning{He et al.}
\titlerunning{NH3/CO2 and the 2152 cm$^{-1}$ CO band}

\abstract
 {CO is one of the most abundant ice components on interstellar dust grains. When it is mixed with amorphous solid water (ASW) or located on its surface, an absorption band of CO at 2152 cm$^{-1}$ is always present in laboratory measurements. This spectral feature is attributed to the interaction of CO with dangling-OH bonds (dOH) in ASW. However, this band is absent in observational spectra of interstellar ices. This raises the question whether CO  forms a relatively pure layer on top of ASW or is in close contact with ASW, but not via dangling bonds. }
 {We aim to determine whether the incorporation of NH$_3$ or CO$_2$ into ASW blocks the dOH and therefore reduces the 2152 cm$^{-1}$ band. }
 {We performed laboratory experiments to simulate the layered structure of the ice mantle, that is, we grew CO ice on top of 1) pure ASW, 2) NH$_3$:H$_2$O=10:100 mixed ice, and 3) CO$_2$:H$_2$O=20:100 mixed ice. Infrared spectra were measured to quantify the strength of the 2152 cm$^{-1}$ band. In addition, a second set of experiments were performed to determine how the incorporation of NH$_3$ into ASW affects the dOH band. }
 {We found that annealing the ice reduces the 2152 cm$^{-1}$ band and that NH$_3$ blocks the dOH on ASW surface and therefore reduces the 2152 cm$^{-1}$ band more effectively than CO$_2$. We suggest that this difference between NH$_3$ and CO$_2$ can be ascribed to the polarity of the guest molecule (NH$_3$ is a polar species, whereas CO$_2$ is apolar). The polarity implies that the formation of an H-bond between the N atom of ammonia and the dOH is a barrier-less reaction. We also determined the pore surface area of the ice mixtures as a function of the annealing temperature, and found that the nondetection of 2152 cm$^{-1}$ band does not necessarily exclude the possibility of a porous ice mantle. }
 {}


 \maketitle

\section{Introduction} \label{sec:intro}
The nondetection of the 2152~cm$^{-1}$ (4.647~$\mu$m) band in astronomical observations of interstellar ices is one of the long-standing puzzles of astrochemistry. The 2152~cm$^{-1}$ band is the result of the interaction between absorbed CO and dangling-OH bonds (dOH) on amorphous water-ice surfaces \citep{Sandford1988, Buch1991, Devlin1992, Palumbo1992, Palumbo1997, Collings2003a, Al-Halabi2004, Palumbo2010, He2019asw}.
This feature has been identified in laboratory CO:\ce{H2O} ice mixtures, whereas it appears to be absent in observational spectra of interstellar ices.  The question then is why CO molecules apparently do not bind to the dangling-OH site in pre- and protostellar environments, resulting in spectra lacking the 2152~cm$^{-1}$ band.

At first, the nondetection was attributed to the low signal-to-noise ratio of observations in the 4 $\mu$m region, and thus, to the difficulties in resolving this band when the abundance of the responsible interstellar molecules is too low (e.g., \citealt{Sandford1988,Ehrenfreund1997}). A second explanation ascribed the nondetection to the low spectral resolution of the observations because of the overlap of several bands in the 4 $\mu$m region, such as gas-phase CO rovibrational transitions, the H$_2$ Pfund-$\beta$ emission line at $\sim$2149~cm$^{-1}$ \citep{Pontoppidan2002}, and the C$-$N stretching mode at 2165~cm$^{-1}$ \citep{Schmitt1989,Shutte1997,Pendleton1999,Novozamsky2001}.  

With the advent of higher-resolution ground-based infrared instrumentation, it has been possible to disentangle the spectral features in the 4 $\mu$m region. In particular, the CO ice band was successfully resolved toward a sample of 39 low- to high-mass young stellar objects (YSOs) using the Very Large Telescope Infrared Spectrometer And Array Camera (VLT-ISAAC; \citealt{Pontoppidan2003}). In the VLT-ISAAC $3-5~\mu$m survey the 2152~cm$^{-1}$ feature was not detected toward any source, and only stringent upper limits were provided \citep{Pontoppidan2003}. The large sample size (i.e., 39 objects), high signal-to-noise ratio, and spectral resolution of this set of observations suggested that the nondetection of the 2152~cm$^{-1}$ band is not linked to the capabilities of existing infrared facilities, but might be linked to the structure and composition of interstellar ices. It is well established that water-based ices in the interstellar medium (ISM) are amorphous, and a good analog of them is amorphous solid water (ASW). ASW grown by vapor deposition at low temperatures ($\sim$10$-$20~K) in the laboratory is highly porous with a very large specific surface area \citep{Stevenson1999, Kimmel2001, Ayotte2001, He2019asw}. As the temperature becomes higher, the porosity decreases, and the crystallinity increases. The blue wing of the broad OH-stretching absorption peak in the infrared contains two weak dOH absorption features at 3696 and 3720 cm$^{-1}$ \citep{Rowland1991, Buch1991, Palumbo2005,Raut2007Characterization, Palumbo2010, Dartois2013, Bu2015,He2019asw}, which are attributed to 3- and 2-coordinated water sites on the pore surface, respectively. It has been demonstrated that the 3720 cm$^{-1}$ (2.688 $\mu$m) feature is unstable and diminishes when the ASW is slightly annealed or exposed to irradiation \citep[e.g.,][]{Palumbo2005, Raut2007Compaction, Bu2015, He2019asw}, while the 3696 cm$^{-1}$ (2.705 $\mu$m) feature persists even when the ASW is annealed to as high as 130~K. The more stable 3696 cm$^{-1}$ feature was searched for in the ISM, but was not found \citep{Keane2001a}. It is known that the 2152~cm$^{-1}$ feature is linked to the interaction of CO with dOH \citep{Devlin1992}, and that there is an anticorrelation between the 2152~cm$^{-1}$ and 3696~cm$^{-1}$ bands \citep{He2019asw}. Therefore, a theory that explains the nondetection of 2152~cm$^{-1}$ feature when CO is present in the ice should also explain the nondetection of 3696~cm$^{-1}$ when CO is absent. 

Several laboratory studies have been devoted to studying the structure of water ice (porosity, pore surface area, etc.) and the dOH bonds, particularly the impact of energetic particles. \citet{Raut2007Compaction} used 100 keV Ar ions to bombard ASW and found compaction of the ice and a decrease in the dOH peak area. They proposed that impacts from cosmic rays can explain the absence of dOH features in molecular clouds. \citet{Palumbo2010} used keV ions as well as Lyman-alpha photons (UV) to irradiate water-containing ices, and found that both UV and ions reduce the dOH band area and thus help to explain the nondetection of dOH feature. \citet{Dartois2013} used ions of even higher energy (MeV to GeV) to better simulate the effect of cosmic-ray bombardment and confirmed the compaction of ASW and the elimination of dOH bonds by ions. \citet{Behr2020} varied the temperature of ASW between 10 and 90~K and found that the decrease in dOH band area occurs in the whole temperature range they explored. These experimental works mostly focused on the $\sim$3700~cm$^{-1}$ dOH feature, but not directly on the 2152~cm$^{-1}$ feature. All of these experimental studies indicate that energetic particles can make porous ASW compact and reduce the dOH feature. However, the compaction of pure ASW by either thermal processing or bombardment of energetic particles is unlikely to fully eliminate the dOH feature because the dOH feature in pure water ice is directly linked to its (pore) surface area. It has been found that regardless of whether the water ice is amorphous or crystalline, dOH is always present \citep{Zhang1990a, Zhang1990b, Rowland1991, Devlin1992}. The ice mantle might possess a certain level of porosity while at the same time the dOH feature is negligible. The dOH bonds might also be blocked by other molecules. We try to address these issues in the current study. 

The answer to the nondetection of the 2152~cm$^{-1}$ band was also directly investigated with laboratory experiments. \citet{Palumbo1997} measured the infrared spectra of \ce{H2O}:\ce{CO} ice mixtures of different mixing ratios and compared the area of the two peaks at 2139 and 2152~cm$^{-1}$. She found that as the concentration of CO decreases, the 2152~cm$^{-1}$ is more pronounced, suggesting that CO favors the dOH binding sites over other sites. \citet{Fraser2004} carried out experiments of layered CO:\ce{H2O} ices with different CO coverage. The authors inferred that layered CO:\ce{H2O} ices, with significant CO, are good interstellar-ice analogs over mixed ices because the 2152~cm$^{-1}$ feature is not evident in their spectra. This is explained by the fact that CO can only react with a small surface area, resulting in a 2152~cm$^{-1}$ band under the detection limit. In this case, the ASW on dust grains would have a relatively compact structure, and CO would easily form a pure layer on top of the water-dominated layer. However, it is unclear whether the water-dominated layer is fully compact. The absence of 2152~cm$^{-1}$ peak is unlikely to be explained by a compact ASW structure alone. For example, \citet{He2019asw} showed (see Fig. 9 in that paper) that CO on an ASW that was annealed to 140~K and cooled down to 20~K still has the 2152~cm$^{-1}$ peak. For CO on pure ASW, the 2152~cm$^{-1}$ peak is proportional to the surface area of the ASW \citep{He2019asw}. Even when we assume that all the interstellar water ice is fully compact, the surface area could still be significant, considering that dust grains could be fluffy. As an alternative explanation, \citet{Fraser2004} further speculated that some species other than CO, in particular, CO$_2$, NH$_3$, CH$_4$, and CH$_3$OH, could adsorb at dangling-OH binding sites, thus removing the 2152~cm$^{-1}$ feature in observational spectra. In this scenario, the ASW is still allowed to have a large surface area. However, \citet{Fraser2004} did not provide experimental confirmation of which interstellar ice components can act as effective dOH blockers. 

\citet{Cuppen2011} tested this hypothesis by performing a spectroscopic study of CO:H$_2$O:CO$_2$ and CH$_3$OH:CO mixtures. 
They found that a CH$_3$OH:CO 1:1 mixture is sufficient to reproduce the observed width and peak position of the red component of the CO band, as well as the nondetection of the 2152~cm$^{-1}$ band. In this regard, CH$_3$OH:CO mixtures provided a better fit of one low-mass and one high-mass protostars over a H$_2$O:CO mixture \citep{Penteado2015}. This finding suggested that CO might reside in H$_2$O-poor/CH$_3$OH-rich environments in interstellar ices; this concept is supported by laboratory experiments and model simulations of CH$_3$OH formation from CO ice \citep{Watanabe2002,Fuchs2009,Cuppen2009}. 

Although convincing, a mixing ratio of CH$_3$OH:CO 1:1 does not reflect the average composition of interstellar ices \citep{boogert2015}. The CH$_3$OH:CO ratio in high-mass YSOs is in fact subjected to large variations (0.2$-$3; \citealt{Dartois1999,Boogert2002,Boogert2008,Pontoppidan2003,Pontoppidan2008}). In addition, for fewer than a handful of low-mass YSOs, the CH$_3$OH:CO ratio is $\ge 0.7$ \citep{Pontoppidan2004,Perotti2020,Perotti2021}, and it is only 0.2$-$0.3 for starless and prestellar cores \citep{Boogert2011,Chu2020,Goto2021}.  

Additionally, although it is plausible that a fraction of CO settles in CH$_3$OH-rich ices, a significant amount of it inevitably migrates and interacts with the H$_2$O matrix \citep{Collings2003b,He2018diff,Zamirri2018}, contributing to the 2152~cm$^{-1}$ band. This consideration is not taken into account in \citet{Cuppen2011} due to the selection of a CH$_3$OH:CO mixture that does not contain H$_2$O. In summary, although large quantities of CH$_3$OH can act as a dangling-OH bond blocker, other more abundant species present in the ices, such as NH$_3$ and CO$_2$, might also play a role in the suppression of the 2152~cm$^{-1}$ band.

According to mid-infrared observations, the abundance of NH$_3$ ice in low- and high-mass star-forming regions is $\sim 5-7\%$ relative to H$_2$O \citep{Dartois2002,Bottinelli2010,Oberg2011a}. However, laboratory constraints on optical constants and band strengths have shown that its abundance has been underestimated by up to 30\% \citep{Kerkhof1999, Zanchet2013}. NH$_3$ is thought to form on the ice surface by the H-addition to N atoms \citep{Hiraoka1995} while competing with the H-addition to O atoms. This suggests that NH$_3$ is formed simultaneously with water and is intimately mixed with water. This might also result in newly formed NH$_3$ molecules reacting with dOH. Similarly, the abundance of CO$_2$ ice in the ISM has been estimated to between $10-34\%$ compared to H$_2$O \citep{dhendecourt1989,degraw1996,Gibb2004,Pontoppidan2008,Noble2013}. The most efficient formation pathways of CO$_2$ are proposed to be the reaction between CO and OH radicals \citep{Noble2011, Oba2011, Ioppolo2011} or between CO and oxygen atoms \citep{Roser2001}. Because water is formed from O and OH as well, it follows that CO$_2$ should also be intimately mixed with water. It is possible that CO is binding to the dOH sites before forming CO$_2$, and therefore blocking the dOH sites \citep{Fraser2004}. 

A related question that must be addressed is the quantification of the surface area of ices in the ISM. Ices play an important role in chemical reactions in the ISM because they provide a surface to adsorb and retain small reactive species, and because they stabilize reaction products by taking away the energy released in chemical reactions. The reaction rate depends on whether reactions occur in bulk ice or on the ice surface. The diffusion rate, which largely determines the rate of chemical reactions, is usually much faster on the surface than in bulk. When the chemistry in the ISM is modeled, it is necessary to distinguish surface reactions from reactions in bulk ice. For this reason, laboratory measurement of the surface area of astrophysically relevant ices provides key information for an accurate modeling of grain-surface reactions. As demonstrated by \citet{He2019asw}, the pore surface area of ASW decreases with increasing annealing temperature, and the surface area is proportional to the band area of the 3696~cm$^{-1}$ dOH feature. It is therefore important to examine the pore surface area together with the dOH feature at 3696~cm$^{-1}$ or with the 2152~cm$^{-1}$ feature. 

In the present paper, we test whether the absence of the 2152 cm$^{-1}$ feature is due to the interaction of CO with water ice and whether molecules such as NH$_3$ and CO$_2$ could block the dOH signatures of ASW. We performed a systematic study of different H$_2$O:NH$_3$ and H$_2$O:CO$_2$ ice mixtures, using porous ASW as the main ice analog component. CO was deposited on top of both mixtures. NH$_3$ and CO$_2$ were selected because they are ubiquitous species in astrophysical environments. In particular, they have been detected in the gas reservoirs of inner and outer regions of protoplanetary disks and in protostellar cores \citep{Guertler2002, Gibb2004, Oberg2010, Mandell2012, Salinas2016, Bosman2017, Pontoppidan2019, Najita2021} and on comets and planets \citep{Mumma2011, Altwegg2020, Swain2008, Millan2021}. This work focuses on the 2152 cm$^{-1}$ feature; we do not attempt to tackle the red component at $\sim$2136 cm$^{-1}$. We do not compare the measured spectra with observations either because our measured spectra are in reflection mode, which should not be compared directly with observations. 

This paper is structured as follows. We begin in Section~\ref{sec:experimental setup} by describing the experimental setup used to record the laboratory spectra. In Section~\ref{sec:experimental results and analysis} we test the effectiveness of \ce{CO2} and \ce{NH3} as dOH bond blockers and present the key experimental results. Section~\ref{discussion} discusses the results and explains them by comparing the chemical behaviors of CO$_2$ and NH$_3$ embedded in H$_2$O. Additionally, it supplies a list of additional dOH-blocker candidates. We summarize our conclusions in Section~\ref{sec:conclusions}.

\begin{figure*}
    \includegraphics[width=0.99\linewidth]{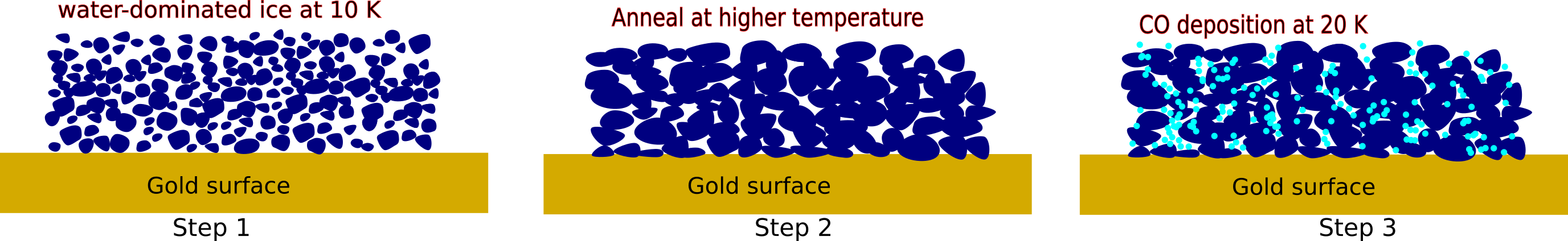}
\caption{Cartoon of the procedure of the first set of experiments. Step 1: Water-dominated ice was deposited at 10 K, forming a highly porous amorphous ice. Three different ice compositions were chosen: (1) 200 ML \ce{H2O}, (2) mixture of 200 ML \ce{H2O} and 20 ML \ce{NH3}, and (3) mixture of 200 ML \ce{H2O} and 40 ML \ce{CO2}. Step 2: Ice was warmed up to a higher temperature to anneal for 30 minutes. The following temperatures were chosen: 20, 40, 60, 80, 100, or 120 K. Step 3: Ice was cooled down to 20 K and exposed to CO deposition. RAIRS spectra were recorded during the CO deposition, which is shown in Fig.~\ref{fig:spectra}. }
 \label{fig:exp_procedure}
\end{figure*}

\section{Experimental setup}
\label{sec:experimental setup}
Experiments were performed using an ultra-high vacuum (UHV) chamber that has been described previously \citep{He2018co2,He2018diff,He2019asw}. The main UHV chamber was pumped to a base pressure of $4\times10^{-10}$ Torr when the cryostat was off. A gold-coated copper disk was used as the substrate onto which ice samples were grown. The temperature of the substrate can be controlled between 5 and 400 K by using a closed-cycle helium cryostat and a cartridge heater installed right underneath the substrate. The temperature was measured using a calibrated silicon diode sensor to an accuracy of 0.05 K. Gas or vapor was deposited through one of the two UHV variable leak valves, which were controlled by stepper motors interfaced to a computer. A LabVIEW program automates the gas and vapor deposition and calculates the thickness in real-time. A more detailed description of the gas deposition is given in the Appendix of \citet{He2018co2}. Although the UHV system is equipped with two highly collimated molecular beam lines for gas deposition, we only used the leak valves because the background deposition from leak valves is a better simulation of the omnidirectional deposition of gases in interstellar space. Water vapor was evaporated from distilled water, which underwent at least three freeze-pump-thaw cycles to remove dissolved air.
%The purity for CO and \ce{NH3} are 99.9\% and 99.99\%, respectively. 
Ice samples were analyzed by a Nicolet 6700 Fourier Transform InfraRed (FTIR) Spectrometer in the reflection absorption infrared spectroscopy (RAIRS) configuration. Infrared spectra were obtained by averaging eight scans every 10 seconds in the range 650--4000 cm$^{-1}$. The infrared beam is at a 78 \textdegree incidence angle relative to the surface normal. In this reflection mode, the infrared band profile may differ from the transmission mode, such as that in \citet{Cuppen2011}. Particularly, the longitudinal optical (LO) mode of CO at 2143~cm$^{-1}$, which appears when the coverage of CO on the surface is relatively high, is only present in reflection mode. In this study, we mostly focus on the relatively weak dOH features at low coverage of CO. Here the difference between reflection and transmission geometry is not important. 

\section{Experimental results and analysis}
\label{sec:experimental results and analysis}
\subsection{CO deposition on water-dominated ice mixtures}
\label{sec:co_on_mixture}
\begin{figure*}
\centering
    \includegraphics[width=0.9\linewidth]{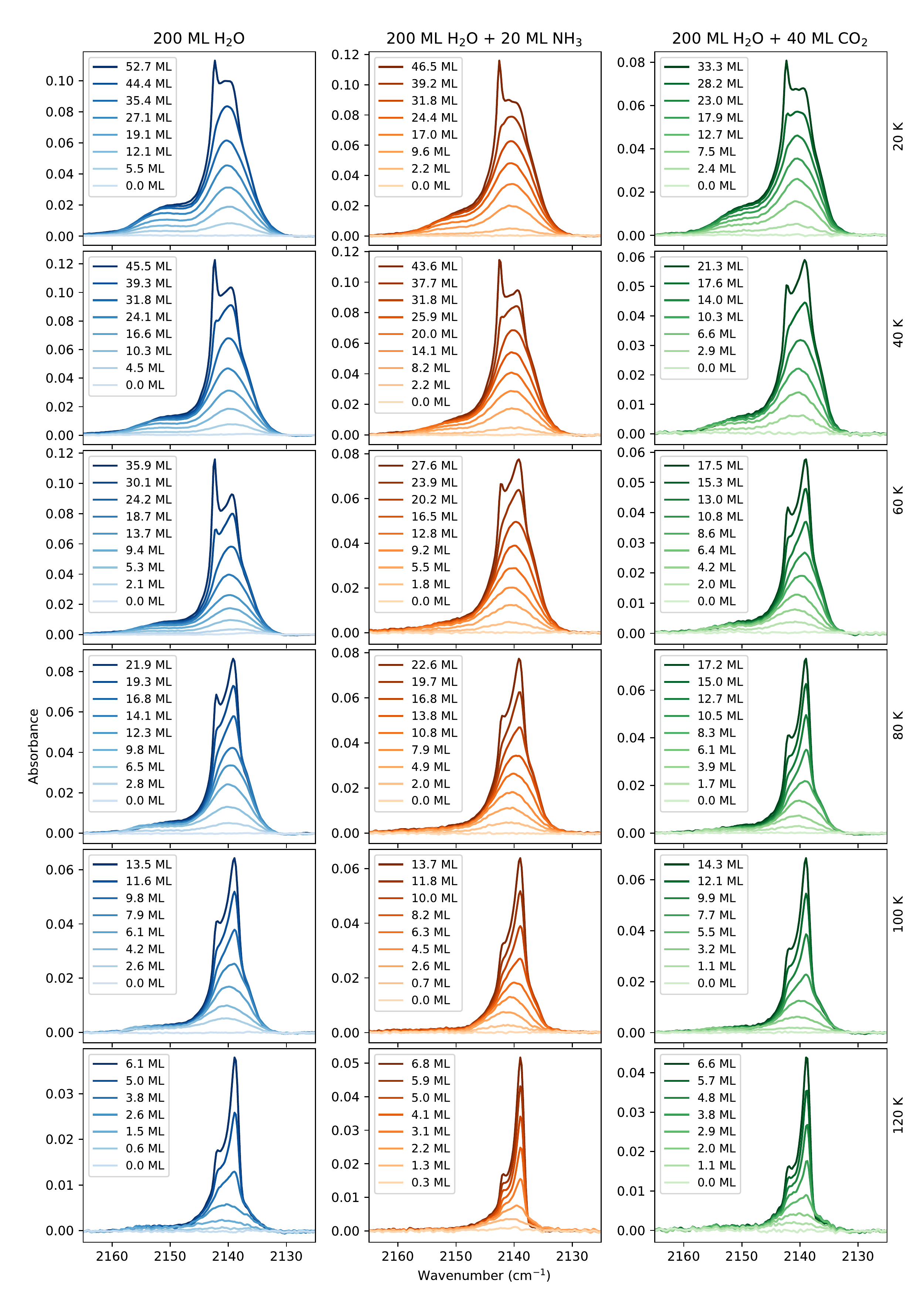}
\caption{Selected RAIRS spectra of CO deposited on water-dominated ice after the ice was annealed to temperature $T_\mathrm{ann}$ and then cooled to 20~K. $T_\mathrm{ann}$ is 20, 40, 60, 80, 100, and 120 K from top to bottom. The three columns are for (1) 200 ML of pure H$_2$O ices, (2) mixed ices of 200 ML H$_2$O and 20 ML NH$_3$, and (3) mixed ices of 200 ML H$_2$O and 40 ML CO$_2$. The insets of the figure refer to the amount of deposited CO.}
\label{fig:spectra}
\end{figure*}

Two sets of experiments were performed in the current study; they are described in this subsection and Sect.~\ref{sec:h2o_nh3_mix}. The first set is an extension to the experiments reported in \citet{He2019asw}. \citet{He2019asw} grew 200 monolayers (ML; defined as $10^{15}$ molecule$\cdot$cm$^{-2}$) of water by vapor condensation on the gold surface with the surface at 10 K. At this temperature, the water ice is amorphous and highly porous. Afterward, the water ice was warmed up to an annealing temperature $T_{\rm ann}$ and remained at this temperature for 30 minutes to stabilize the ice structure. The following annealing temperatures were chosen: $T_{\rm ann}$ =  20, 40, 60, 80, 100, and 120~K. After annealing, the ice was cooled down to 20~K before depositing CO on top of the water ice, and RAIRS spectra were collected during the CO deposition. We chose to make a layered ice rather than a uniformly mixed one to better represent the layered ice mantle on dust grains \citep{Pontoppidan2008, boogert2015}. This is different from other studies, in which gases were premixed and then deposited \citep[e.g.,][]{Ehrenfreund1997,Cuppen2011}. In addition to the experiments in \citet{He2019asw}, in the present study, we replaced pure water with (1) mixed ice of 200 ML \ce{H2O} and 20 ML \ce{NH3} or (2) mixed ice of 200 ML \ce{H2O} and 40 ML \ce{CO2}. The ratios \ce{CO2}:\ce{H2O} and \ce{NH3}:\ce{H2O} are representative of that in the ice mantle on dust grains \citep{Pontoppidan2008, Oberg2011a, Zanchet2013, boogert2015}. Other than the difference in ice composition, the experiments follow the same procedure as in \citet{He2019asw}. The deposition of \ce{H2O}:\ce{CO2} and \ce{H2O}:\ce{NH3} ice mixtures was accomplished by codeposition of the two gas and vapor components within 20 minutes and 25 minutes, respectively. Fig.~\ref{fig:exp_procedure} demonstrates the procedure of the experiments. RAIRS spectra were measured during CO deposition on ices, a selection of which is shown in Fig.~\ref{fig:spectra}.

\begin{figure}
\centering
    \includegraphics[width=0.99\linewidth]{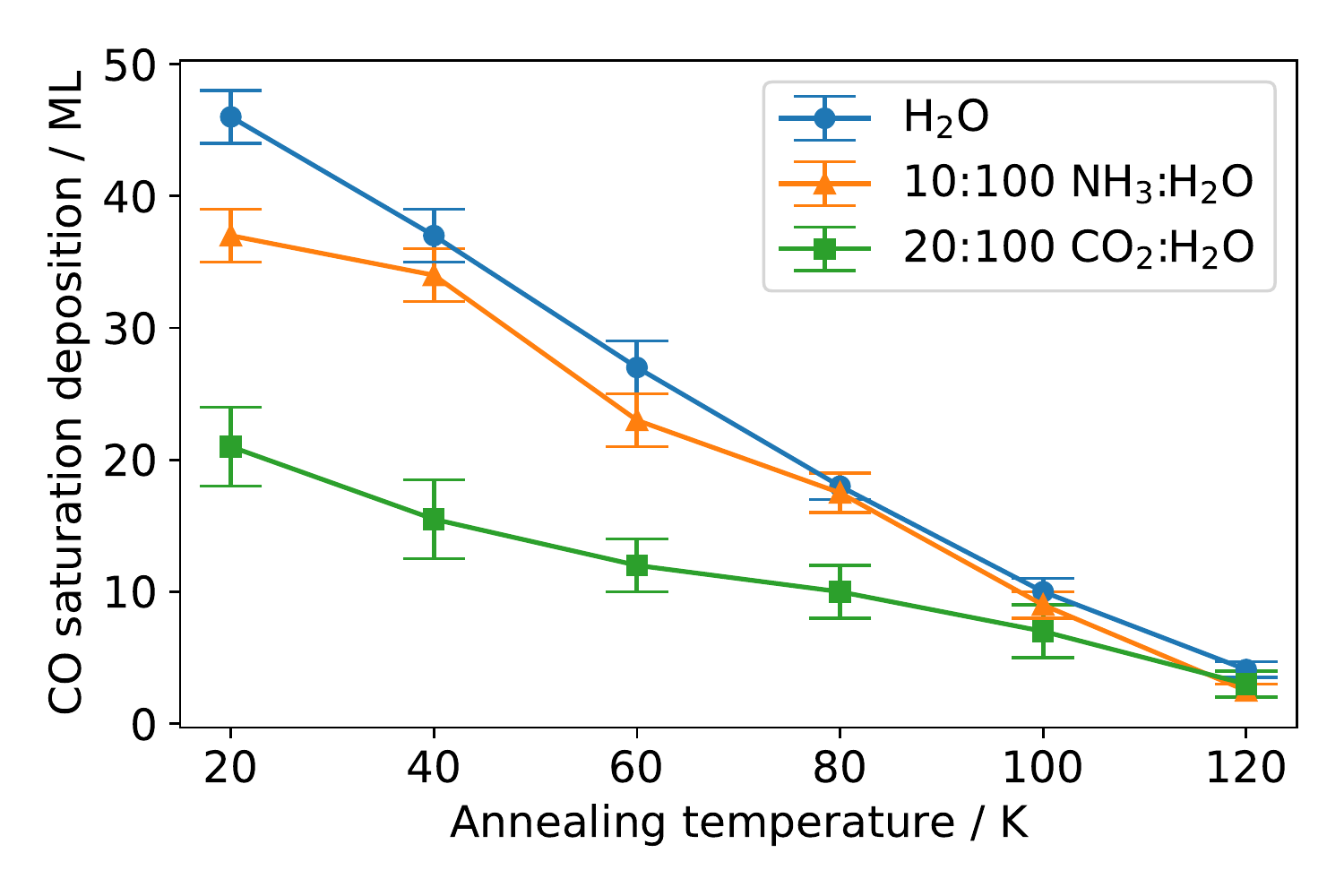}
\caption{Pore surface area of water-dominated ice (mixture) that is annealed to different temperatures, as determined from Fig.~\ref{fig:spectra}. The surface area is the CO deposition dose at which the 2143 cm$^{-1}$ peak emerges. }
\label{fig:sat_area}
\end{figure}

It has been shown that CO molecules can diffuse and penetrate porous ASW already at below 16-20~K, reaching all the surface area pores \citep{Mispelaer2013, Karssemeijer2014, Lauck2015, He2018diff, He2019asw}. 
The same is true in the \ce{H2O}:\ce{CO2} and \ce{H2O}:\ce{NH3} mixtures. The second and third columns of Fig. \ref{fig:spectra} show that when the CO coverage is below saturation of the surface, the peak profile is similar to the profile that is usually seen in CO and \ce{H2O} mixed ice. At the beginning of the CO deposition, the coverage of CO on the pore surface is small, and all the CO molecules interact with the water surface, which is shown by the two peaks in the infrared centered at $\sim$2139 and $\sim$2152 cm$^{-1}$. 
As CO gas deposition proceeds, the pore surface area is gradually covered by CO, and eventually, CO builds up as pure CO ice; at this stage, CO primarily interacts with other CO molecules rather than with the water surface. This is indicated by the emergence of a 2143 cm$^{-1}$ absorption peak due to the LO mode of CO. Similar to \citet{He2019asw}, we took the CO deposition amount at which the 2143~cm$^{-1}$ peak emerges as the pore surface area of the ASW. Fig.~\ref{fig:sat_area} shows the pore surface area, in units of ML, of the three ices at different annealing temperatures. The error bar accounts for the uncertainty in determining the emergence of the LO peak. 
The introduction of 20\% \ce{CO2} into water ice lowers the pore surface area by almost half, while 10\% of \ce{NH3} has only a small impact on the surface area. One possible reason why \ce{CO2} more effectively blocks the pore surface area is that \ce{CO2} forms clusters on the surface of ASW \citep{He2017CO2}, which block some micropores.

\begin{figure*}
\centering
    \includegraphics[width=0.99\linewidth]{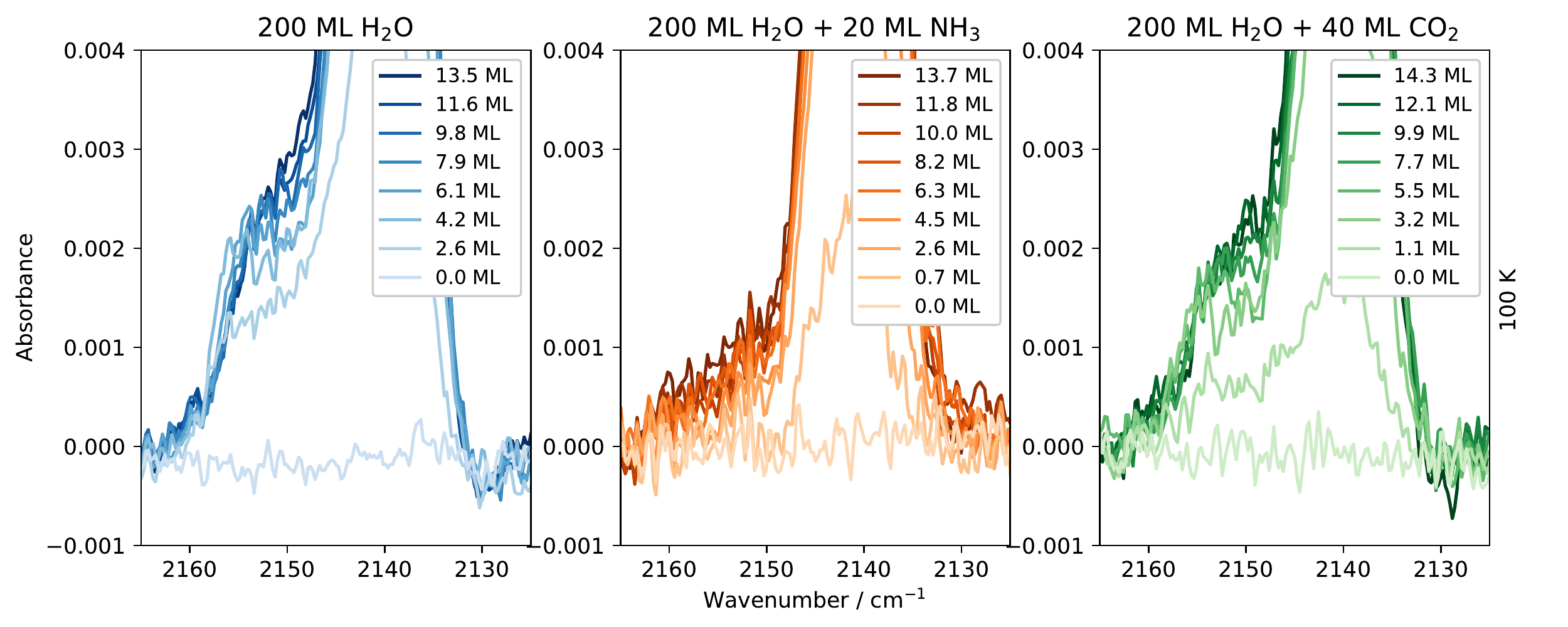}
\caption{Zoom of the fifth row (100~K) of Fig.~\ref{fig:spectra}.  }
\label{fig:spectra_zoom}
\end{figure*}

The band area of the 2152 cm$^{-1}$ peak in Fig.~\ref{fig:spectra} can also be analyzed. By visual inspection, we can see that the peak area decreases at higher annealing temperatures for all three ices. When annealed at 120 K, the 2152 cm$^{-1}$ peak becomes insignificant. In \ce{H2O}:\ce{NH3} mixtures, the decrease in 2152 cm$^{-1}$ peak area is more pronounced. Even if the annealing temperature is only  60~K, the 2152~cm$^{-1}$ peak appears to be the tail of the 2143~cm$^{-1}$ peak rather than a separate peak. When the \ce{H2O}:\ce{NH3} mixture is annealed to 100~K, the 2152 cm$^{-1}$ peak disappears almost completely, while in the \ce{H2O}:\ce{CO2} mixture, the 2152 cm$^{-1}$ peak is still significant. Fig.~\ref{fig:spectra_zoom} shows a zoom-in of the fifth row of Fig.~\ref{fig:spectra}, corresponding to the ices annealed to 100 K.  Because the 2152 cm$^{-1}$ peak is directly correlated to CO interacting with the dOH bonds \citep{He2019asw}, it is evident that 10\% of \ce{NH3} blocks the dOH bonds more effectively than 20\% of \ce{CO2}, provided that the ice is annealed to higher than 80~K. As discussed above, 20\% of \ce{CO2} reduces the surface area of ASW by half, which should also reduce the dOH sites proportionally. In the \ce{CO2}:\ce{H2O} mixture we note that the surface area decreases to half throughout the experiment (Fig.~\ref{fig:sat_area}), but in contrast to expectations, the 2152 cm$^{-1}$ peak is not equally reduced to half. This observation strengthens the argument that \ce{CO2} is a less efficient dOH blocker.

\begin{figure}
\includegraphics[width=0.99\linewidth]{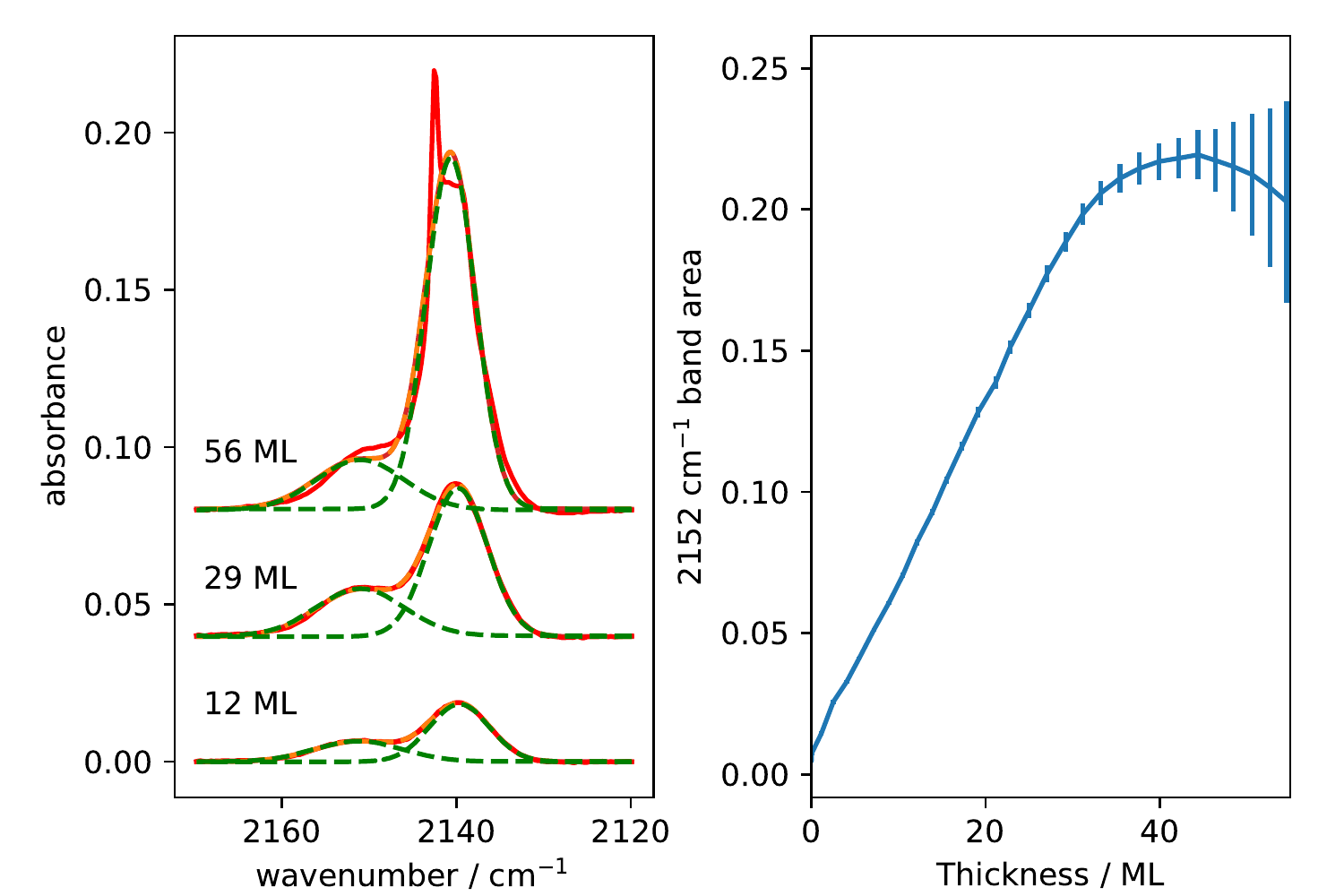}
\caption{Illustration of the method for determination of 2152~cm$^{-1}$ band area. 
Left panel: Examples of the fitting of the CO absorption profile using the sum of two Gaussian functions. The red lines are the measured spectra, the dashed green lines are the two Gaussian components, and the orange lines are the total fitting. Spectra are offset for clarity. Right panel: Band area of the 2152~cm$^{-1}$ component during CO deposition; the error bar of the fitting is also shown. The large error bar at larger thickness is due to the emergence of the LO peak at 2143~cm$^{-1}$.  \label{fig:fit_example}}
\end{figure}

\begin{figure}
\centering
\includegraphics[width=0.9\linewidth]{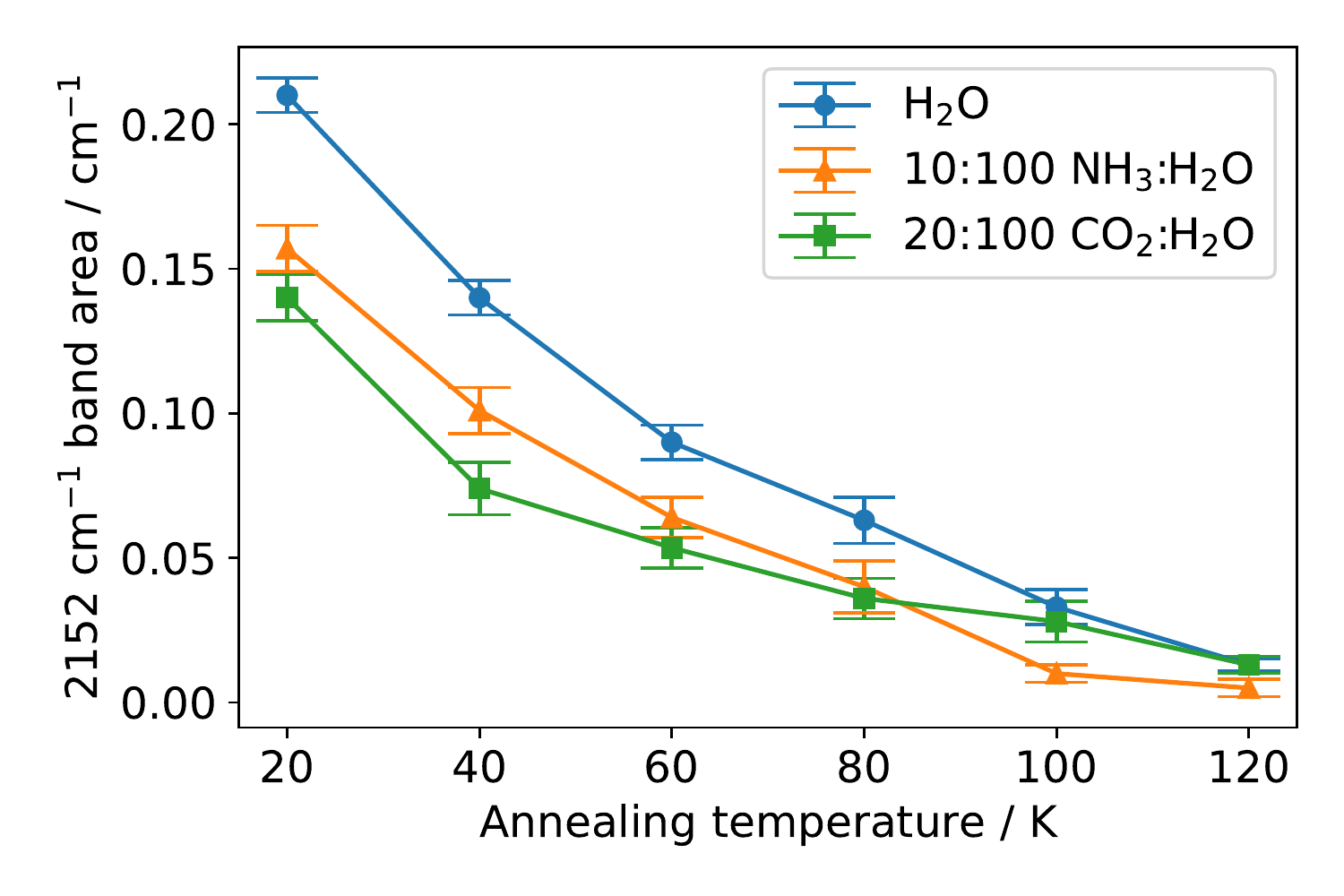}
\caption{Saturation band area for the 2152 cm$^{-1}$ band, as shown in Fig.~\ref{fig:spectra}. The band area is calculated by applying the fitting protocol illustrated in Fig.~\ref{fig:fit_example}.}

 \label{fig:sat_doh}
\end{figure}

To quantify it, we calculated the 2152 cm$^{-1}$ peak area by using the sum of two Gaussian functions to fit the spectra in Fig.~\ref{fig:spectra}. Because the 2143 cm$^{-1}$ peak only emerges after the surface is saturated by CO, that is, the 2152 cm$^{-1}$ peak is saturated before the emergence of the 2143 cm$^{-1}$ peak, we only focused on the spectra before the emergence of the 2143 cm$^{-1}$ peak. The advantage is that we only need two components for the fitting.
Fig.~\ref{fig:fit_example} shows examples of the fitting. With a small amount of CO on the surface, the fitting by the sum of two Gaussian functions is excellent. However, after CO fully covers the surface, the fitting becomes worse, as illustrated by the top trace in the left panel. This is also demonstrated in the right panel. As the surface is fully covered by CO, the error bar of the 2152 cm$^{-1}$ band area becomes very large. From the right panel, we can determine the saturation 2152~cm$^{-1}$ peak areas. This analysis procedure is repeated for the ices with different annealing temperatures, and the results are shown in Fig.~\ref{fig:sat_doh}. Here we can see that 10\% of \ce{NH3} effectively blocks the dangling OH bonds as long as the ice mixture is annealed to ~100~K. 
We note that the fitting only calculates the area of the 2152~cm$^{-1}$ peak, but does not consider its shape. For \ce{H2O}:\ce{NH3} mixtures annealed to 60~K and above, the 2152~cm$^{-1}$ component looks as if it is the tail of the 2140/2143~cm$^{-1}$ peak rather than a separate peak, making it more difficult to discern than for \ce{H2O}:\ce{CO2} mixtures in astronomical observations. 

\begin{figure}
\centering
\includegraphics[width=0.9\linewidth]{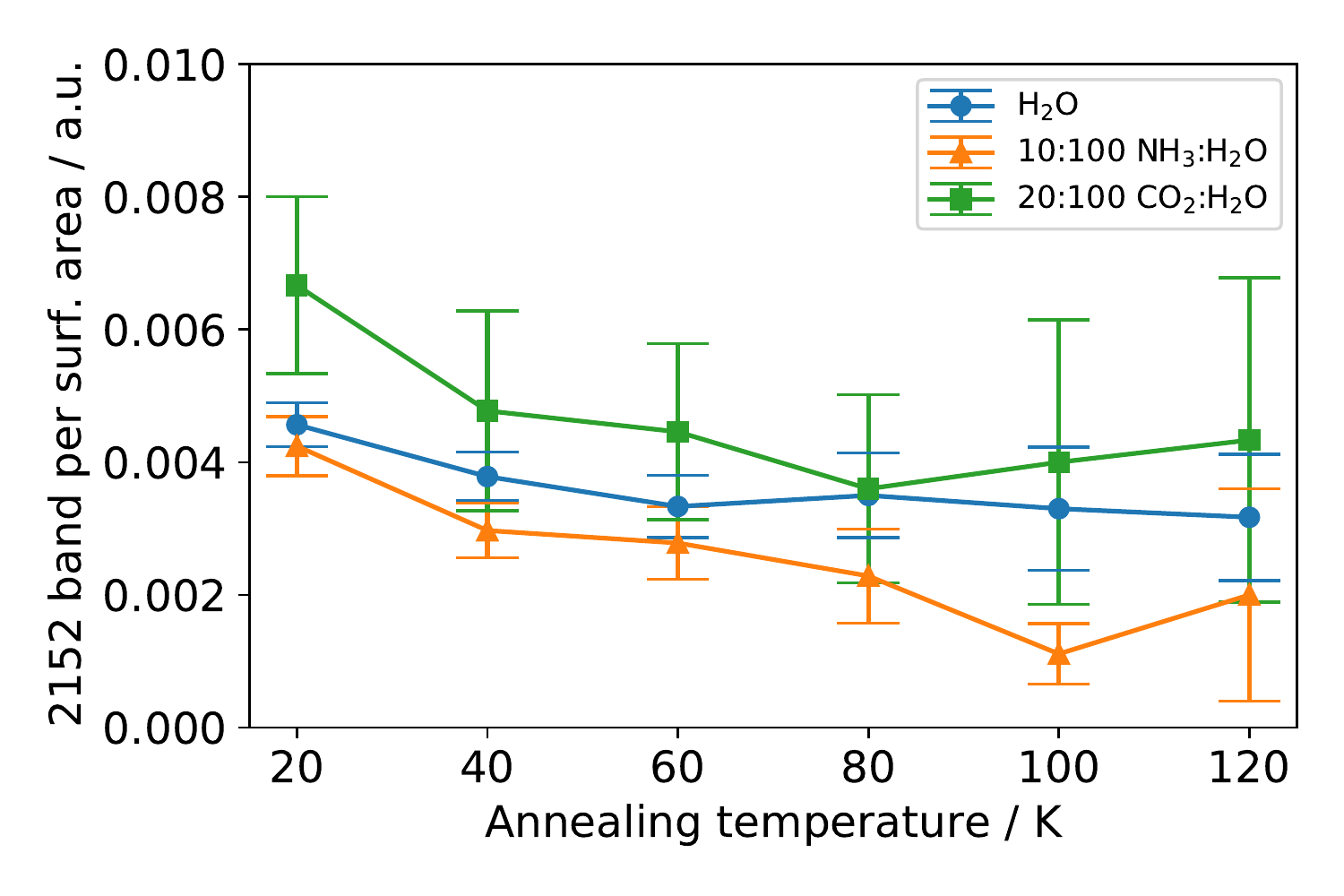}
\caption{Saturation band area for the 2152 cm$^{-1}$ band as shown in Fig.~\ref{fig:sat_doh}, divided by the pore surface area, as shown in Fig.~\ref{fig:sat_area}.}
 \label{fig:2152perSurfArea}
\end{figure}

Introducing \ce{CO2}/\ce{NH3} into ASW has an impact on both the pore surface area and the fraction of dOH sites on the pore surface. It is necessary to determine whether the decrease of dOH is due to the blocking by \ce{CO2}/\ce{NH3} or due to the decrease of the pore surface area. For this reason, the saturation band area for the 2152 cm$^{-1}$ band in Fig.~\ref{fig:sat_doh} was divided by the pore surface area in Fig.~\ref{fig:sat_area} to obtain the normalized dOH blocking efficiency. The result is shown in Fig.~\ref{fig:2152perSurfArea}. \ce{NH3} lowers the 2152 cm$^{-1}$ band area per pore surface area compared with pure ASW. In contrast, \ce{CO2} even increases the 2152 band area per pore surface area because it reduces the pore surface area more effectively. This confirms that \ce{NH3} is an effective dOH blocker on ASW surface.

\subsection{Dangling-OH bonds in \texorpdfstring{\ce{H2O}:\ce{NH3}}{H2O:NH3} ice mixtures}
\label{sec:h2o_nh3_mix}
The results of the first set of experiments presented above suggest that introducing \ce{NH3} into water-dominated ice can effectively decrease the 2152 cm$^{-1}$ peak and that \ce{NH3} is an effective dOH bond blocker. In the second set of experiments, we further explore how the concentration of \ce{NH3} and the temperature of the ice affect the dOH bonds.  We fixed the amount of water to be 100 ML, and deposited \ce{NH3}:\ce{H2O} ice mixtures with mixing ratios ranging from 0:100 to 30:100. After the deposition of the mixture at 10 K, the ice was warmed up at a ramp rate of 6~K/minute, and RAIRS spectra were collected.

\begin{figure*}
\centering
\includegraphics[width=0.9\linewidth]{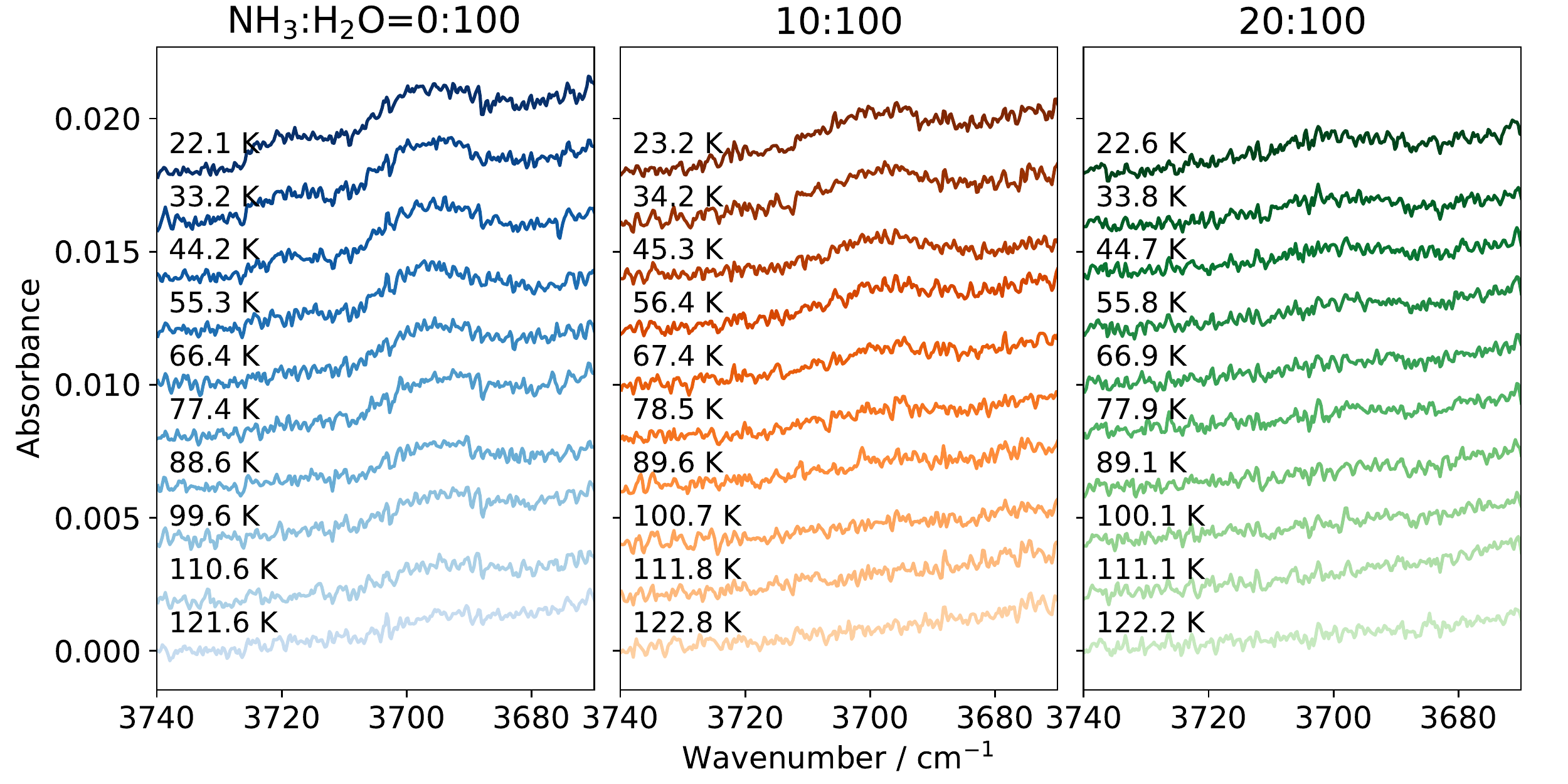}
\caption{Dangling-OH bond region of the RAIRS spectra for \ce{NH3}:\ce{H2O} mixtures during warming up. The mixing ratios and temperatures are indicated in the figure. }  
\label{fig:doh_stack}
\end{figure*}

The spectra of ice with mixing ratio of \ce{NH3}:\ce{H2O}=0:100, 10:100, and 20:100 during warm-up are shown in Fig.~\ref{fig:doh_stack}. In pure water ice, both the dOH bands at 3696 and 3720 cm$^{-1}$ are present at the lowest temperature. The 3720 cm$^{-1}$ band disappears by about 60 K, and the 3696 cm$^{-1}$ band persists until above 120 K. For a more detailed discussion of the dOH bands of pure ASW, see \citet{He2019asw}. When 10\% or 20\% of \ce{NH3} is added to the mixture, the 3720 cm$^{-1}$ peak is indiscernible, except for the 10\% mixture at the lowest temperature. The 3696 cm$^{-1}$ peak is much lower than that for pure ASW, particularly when the ice mixture is warmed up. 

\begin{figure}
\centering
\includegraphics[width=0.9\linewidth]{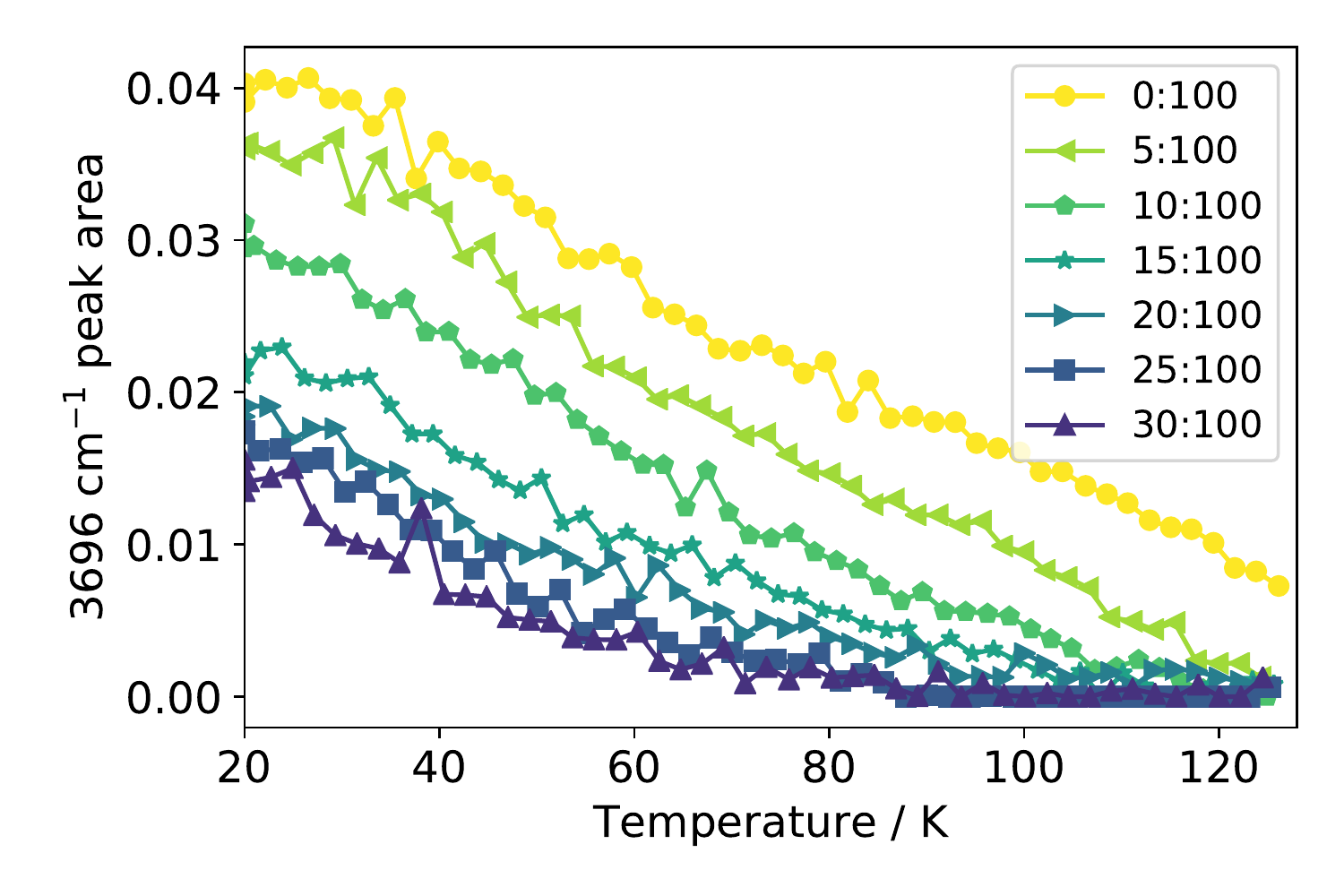}
\caption{Dangling-OH bond  3696 cm$^{-1}$ band area in \ce{NH3}:\ce{H2O} mixtures during warming up. The mixing ratios and temperatures are indicated in the figure. \label{fig:wat_nh3_doh_area} } 
\end{figure}

Following a similar method as in \citet{He2019asw}, we calculated the area for the 3696 cm$^{-1}$ peak. A broad Gaussian function was used to fit the blue wing of the OH-stretch peak, and two narrow Gaussian functions were used to fit the peaks at 3696 and 3720 cm$^{-1}$. The band area of the 3696 cm$^{-1}$ peak was obtained from the fitting. The results are shown in Fig.~\ref{fig:wat_nh3_doh_area}. With an \ce{NH3}:\ce{H2O} mixing ratio between 0 and 15\%, the dOH band area decreases with increasing \ce{NH3} concentration at all temperatures. At higher mixing ratios, the dOH band area becomes less dependent on the concentration. When the mixing ratio is higher than 10\%, the dOH band is almost negligible when annealed to 100 K or above. This agrees with the results of the 2152 cm$^{-1}$ peak in Section \ref{sec:co_on_mixture}. This is expected because it is known that the 2152 cm$^{-1}$ peak is correlated with the dOH band at 3696 cm$^{-1}$ \citep{He2019asw}.

\subsection{\texorpdfstring{\ce{H2O}:\ce{CO2}}{H2O:CO2} ice mixtures}
Similar experiments as in Section~\ref{sec:h2o_nh3_mix} might be performed for CO$_2$:H$_2$O mixtures, and the dOH region might be monitored in the infrared. Such experiments have already been performed and presented in Figure 3 of \citet{He2018co2}. The dOH bonds of water are shadowed by the $\nu_1+\nu_3$ combination mode, and therefore it is challenging to analyze the dOH at 3696 cm$^{-1}$ similar to that in Section~\ref{sec:h2o_nh3_mix}. Previously, \citet{Cuppen2011} measured the infrared spectra of H$_2$O:CO:CO$_2$ mixtures, and found that it takes a large amount of CO$_2$ to suppress the 2152~cm$^{-1}$ band at low temperatures. Even with a mixture of H$_2$O:CO:CO$_2$=2.75:1:10, that is,  three to four times more CO$_2$ than H$_2$O, there is still a non-negligible 2152~cm$^{-1}$ peak. It is clear from both \citet{Cuppen2011} and this study that CO$_2$ is a less effective dOH blocker. 

\section{Discussion}
\label{discussion}

\subsection{The surface area and the nondetection of dOH bands}
The porosity of the ice mantle covering dust grains in the ISM is still debated. Porous water ice provides a large catalytic surface area for grain surface reactions to take place, while reactions in compact ice are challenged by high-diffusion energy barriers for most particles inside bulk compact water ice. The detection of a large variety of complex organic molecules seems to favor the argument that the water-rich ice is somewhat porous and provides a large surface area. However, similarly to the 2152 cm$^{-1}$ band, also the 3696 cm$^{-1}$ dOH band, which is usually present in vapor deposited ASW, has never been detected in interstellar ices \citep{Keane2001a}. This raises the questions why the dOH bands disappear, and whether there is a dOH blocker in the ice. Closely related is the nondetection of the 2152 cm$^{-1}$ peak for CO ice on water surface.  

Previous studies have found that energetic particles bombardment as well as thermal processing reduce the dOH bonds in ASW and make the ice more compact \citep[e.g.,][and references therein]{Raut2007Compaction, Palumbo2006, Dartois2013, He2019asw}. There is also evidence that the ability of CO to diffuse into the ASW is affected by ion processing, which then reduces the 2152 cm$^{-1}$ peak \citep{Palumbo1997}. However, these previous studies did not explore the possibility of having a somewhat porous ice structure without the simultaneous presence of the 2152  cm$^{-1}$  peak (or the 3696 cm$^{-1}$ peak). Our study fills the gap and determines whether dOH might be blocked by other molecules in the ice. We determine whether the ice mantle covering dust grains can have an insignificant 2152 cm$^{-1}$ peak (or equivalently, insignificant 3696 cm$^{-1}$ dOH band), but still have a significant surface area. Figs.~\ref{fig:sat_area}, \ref{fig:sat_doh}, and \ref{fig:wat_nh3_doh_area} show that this is possible if 10\% of \ce{NH3} is mixed with water and the ice is annealed to 100~K (this is on the laboratory timescale, it will be lower on the timescale of a molecular cloud). In the ISM, ices are not condensed from gas and annealed like in the experiments, but are formed by chemical reactions on the surface. The energy released from chemical reactions can be absorbed by the ice and can create some localized ``annealing'', which makes the ice less porous \citep{Oba2009, Accolla2011}. The bombardment of cosmic rays can also induce some local heating of ice and therefore change the ice structure \citep{Palumbo2006, Raut2007Compaction, Dartois2013, Mejia2015, Behr2020}.  Eventually, the state of the ice might be similar to the ice that is condensed from gas and then annealed to a high temperature, such as 100~K.  In this scenario, the 3696 cm$^{-1}$ dOH peak or the 2152 cm$^{-1}$ peak becomes negligible when ammonia is present. We speculate that this may explain the nondetection of these two features.

\begin{figure*}
\centering
\includegraphics[trim={0 0 0 0}, clip,width=0.70\linewidth]{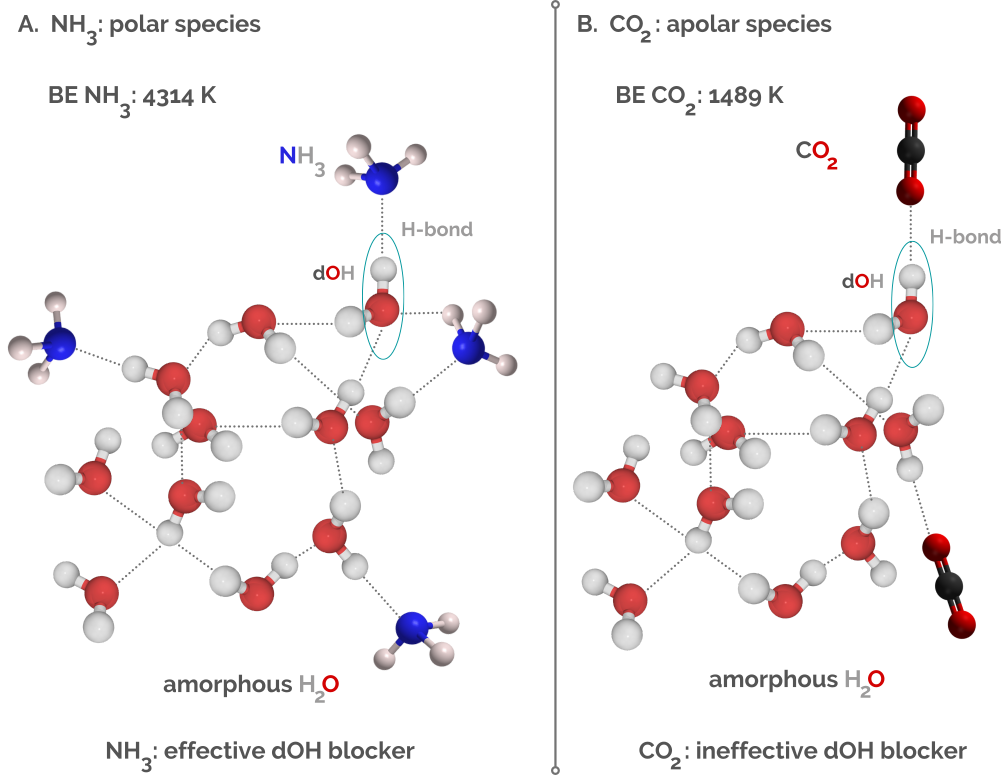}
\caption{Schematic illustration of the interactions between A) dOH and ammonia (NH$_3$) and B) dOH and carbon dioxide (CO$_2$). NH$_3$ is a polar molecule, it tightly binds to the dOH via H-bonding between the N atom and the dOH. CO$_2$ is an apolar species, with binding energy three times lower than that of NH$_3$. This implies that NH$_3$ is a more effective dangling-OH band blocker over \ce{CO2}. The reported binding energies are from \citet{Ferrero2020}. The simplified geometries are an adaptation of real geometries obtained with quantum chemical calculations in \citet{Ferrero2020}.}
\label{fig:cartoon}
\end{figure*} 

\subsection{Effect of polarity on the nondetection of the dOH bands}
The important question that arises after analyzing the different responses of H$_2$O:NH$_3$ and H$_2$O:CO$_2$ ice mixtures is what makes NH$_3$ a more effective dOH blocker than CO$_2$. The answer to this question can be found by comparing the hydrophilicity and the chemical characteristics and behaviors of NH$_3$ and CO$_2$. 

The blocking of dOH bonds on deuterated ASW (d-ASW) has been studied by \citet{Devlin1992}. In this work, d-ASW was precovered with \ce{CF4} or ethylene oxide, and then the infrared spectra of the dOH band at $\sim$2730~cm$^{-1}$ and the band at 2152~cm$^{-1}$ were monitored to investigate the interaction of CO with dOH. \citet{Devlin1992} found that precovering with \ce{CF4} enhances the 2152~cm$^{-1}$ band, whereas precovering with ethylene oxide reduces it. This behavior was attributed to their hydrophilicity. \ce{CF4} is a hydrophobic molecule and therefore avoids dOH bonds and associates strongly with surface oxygen. In contrast, ethylene oxide, as a hydrophilic molecule, primarily binds to surface hydrogen atoms that belong to OH groups, effectively decreasing the number of available dOH groups.

The current study investigates whether \ce{NH3} and \ce{CO2} block the dOH. \ce{NH3} is a hydrophilic molecule such as ethylene oxide, which wets the whole ASW surface before building up as pure layers, as indicated from temperature programmed desorption (TPD) of \ce{NH3} on water ice \citep{He2016binding}. In comparison, \ce{CO2} is hydrophobic, and it forms clusters on ASW surface before the whole surface is covered. The TPDs of \ce{CO2} also show that the interaction force between \ce{CO2} and \ce{CO2} is stronger than that between \ce{CO2} and \ce{H2O} \citep{He2018co2}. This explains why \ce{NH3} blocks dOH while \ce{CO2} does not.

We can also view it from a different aspect by examining their chemical characteristics. NH$_3$ and H$_2$O are structural analogs and isoelectronic species, as they possess the same number of electrons. Both species are nucleophiles: the O atom in H$_2$O and the N atom in NH$_3$ have two unpaired electrons, which can be donated to form hydrogen bonds. Most importantly, because they are characterized by an unequal sharing of valence electrons (i.e., a difference in electronegativity among the bonded atoms), NH$_3$ and H$_2$O are defined as polar. They have permanent dipole moments. In contrast, CO$_2$ is a linear molecule with an equal sharing of valence electrons, hence it is apolar. The fact that NH$_3$ is polar and CO$_2$ is apolar implies that NH$_3$ can bind more tightly to H$_2$O molecules compared to CO$_2$, as confirmed by laboratory measurements using TPD \citep{He2016binding, He2018co2}, which found that a low coverage of NH$_3$ desorbs from water ice at $\sim$140~K, much higher than CO$_2$ , which desorbs at $\sim$80~K. The binding energy on water ice was also studied using quantum chemistry calculations. \citet{Ferrero2020} modeled an interstellar ice surface by simulating both crystalline and amorphous water ices with DFT (B3LYP-D3 and M06-2X). The authors of this work determined binding energies for 21 astrochemically relevant species. On ASW ices, they found that NH$_3$ has binding energy ranges from 4314 to 7549 K, which is several times higher than CO (1109--1869~K) and CO$_2$ (1489--2948~K), corroborating the strong binding between NH$_3$ and water found in experiments. This strong binding makes NH$_3$ a more effective dOH band blocker over \ce{CO2} (Fig.~\ref{fig:cartoon}). 

\citet{Henkelman2016} performed density functional theory (DFT) calculations of the interaction of NH$_3$ with the dOH of water ice and showed that the bond between ammonia and the dOH is promoted by the attraction of the lone pair of the N atom to the dOH of H$_2$O. This is a zero-energy barrier reaction that results in the formation of an OH· · ·\ce{NH3} hydrogen bond. This theoretical study strongly supports laboratory results that proved that NH$_3$ flags the dOH binding sites \citep{Lechner2015}. Both studies revealed that the sole presence of an \ce{NH3} molecule in the vicinity of H$_2$O molecules causes the latter to reorient with zero-energy barrier to expose a dOH bond toward the N atom of \ce{NH3}.  

This is not the case for apolar molecules such as CO or \ce{N2}, which are not capable of binding as strongly as \ce{NH3} to the dOH bonds. Consequently, becausee their attraction to dOH is weaker, their H-bonding with H$_2$O proceeds with nonzero energy barriers. The highest barrier and thus weakest binding to water is found for \ce{N2} , followed by CO  \citep{Henkelman2016}. This implies lower probabilities for the adsorption of \ce{N2} and CO at the dOH sites over \ce{NH3}, and therefore less effective dOH blocking capabilities at the low temperatures of the interstellar medium (ISM). This result agrees with the inefficient dOH blocking behaviors of CO  \citep{Fraser2004} and CO$_2$ (this work and \citealt{Cuppen2011}). 

Previous laboratory studies of the nondetection of the 2152 cm$^{-1}$ band have proposed that good dOH blockers are species that form from CO molecules at the dOH sites (e.g., CO$_2$ and CH$_3$OH; \citealt{Fraser2004,Cuppen2011}; see Section~\ref{other_dOH_blockers}). By working with NH$_3$ and CO$_2$, we have discovered that an important requirement for a good dOH-bond blocker is the polarity of a molecule, its ability to form H-bonds and to reside tightly at the dOH site, therefore impeding CO molecules from reacting with the dOH. As a general rule, we can state that polar molecules, especially NH$_3$ and CH$_3$OH, are efficient dOH blockers, whereas apolar species (e.g., CO$_2$) are not good dOH blockers, unless significantly more abundant than polar species, in agreement with the findings of \citet{Cuppen2011}. Based on the evidence that ammonia is one of the most abundant polar species of interstellar ices after H$_2$O (its abundance is approximately a factor of 3 higher than that of CH$_3$OH; \citealt{Zanchet2013,boogert2015}), we propose that in addition to the impact of thermal annealing and energetic particles, ammonia is one of the main contributors  if probably  not the only one in the suppression of the 2152 cm$^{-1}$ band, and more generally of all the dOH bands.    

\subsection{Other dOH-blocker candidates}
\label{other_dOH_blockers}
The subsequent hydrogenation of N atoms leading to \ce{NH3} at the onset of interstellar ice surfaces is not the only reaction competing with the hydrogenation of O atoms, producing dangling-OH binding sites \citep{Hiraoka1995,Herbst2009}. For instance, the C+H reactions forming \ce{CH4} occur concurrently with the O+H reactions \citep{Qasim2020}. This implies that \ce{CH4} could be a good candidate for the removal or blockage of the 2152 cm$^{-1}$ band binding sites. However, compared to \ce{NH3} and \ce{CO2}, \ce{CH4} is more volatile, with similar desorption characteristics as CO \citep{Collings2004}. Laboratory studies have shown that \ce{CH4} molecules can easily diffuse on the pore surface and do not bind to the ASW surface and hence to the dangling-OH sites tighter than CO \citep{He2018diff}. We further strengthen this argument based on the apolarity of \ce{CH4}, which in turn does not make it an efficient dOH blocker. Other species might act as suppressors of the 2152 cm$^{-1}$ band. Potential candidates are molecules that form onto the ices before the CO condensation stage, which steadily bind to the dOH sites because of their polarity. Examples of this class are OCN and OCS. Finally, one last category of species that could block the dOH sites are polar molecules that are produced by the interaction of CO at the dOH sites with other species present on the ice surfaces. This includes species such as H$_2$CO, HCOOH, and CH$_3$OH or even more complex organic species.

At the current stage, it is unclear whether one species contributes to the suppression of the 2152 cm$^{-1}$ band observed in astronomical spectra more than others, or if this is the result of multiple contributions. Further systematic laboratory experiments of ternary and quaternary mixtures are needed to constrain the composition and structure of ice mantles. In addition, quantum chemical modeling of properly mixed ices, representatives of real astronomical ices (i.e., which include at least CO, \ce{CO2}, \ce{NH3}, \ce{CH4}, and \ce{CH3OH}), may also help to elucidate the structure and spectroscopic features of the CO band, accounting for the interaction of multiple species in different \ce{H2O}–ice structural environments. Finally, high-sensitivity near- and mid-infrared observations of the dOH bands with the \textit{James Webb Space Telescope} will provide spectra with a higher signal-to-noise ratio to compare with model predictions and experimental data.

\section{Conclusions}
\label{sec:conclusions}
We investigated one of the conundrums of astrochemistry: the absence of the 2152 cm$^{-1}$ band in observational spectra of pre- and protostellar environments. This band is due to the interaction between CO and dangling OH bonds on the surface of water ice. By making composite ices in the laboratory that resemble the icy mantles on dust grains, we examined the impact of NH$_3$ and \ce{CO2} on the 2152 cm$^{-1}$ band and the pore surface area of the mixed ice. We found that introducing 10\% of NH$_3$ in ASW effectively blocks the dOH bonds as long as the ice is annealed to 100~K. At the same time, introducing NH$_3$ exerts only a small impact on the pore surface area of ASW. In comparison, introducing 20\% of \ce{CO2} in ASW reduces the pore surface area by half, but does block all the dOH bonds regardless of the annealing temperature. We discussed from a chemistry point of view why polar molecules such as \ce{NH3} are likely more efficient dOH blockers than apolar molecules. We propose that NH$_3$ is one of the most important dOH blockers, which helps to explain the nondetection of the 2152 cm$^{-1}$ band, and the water-dominated layer of the ice mantle can be slightly porous, while the 2152 cm$^{-1}$ peak is insignificant.

\section{Acknowledgements}
The authors wish to thank Prof. Piero Ugliengo for the fruitful discussion on the binding energies of ammonia and carbon dioxide in interstellar ices. J.H., G.P. and T.H. acknowledge support from the European Research Council under the Horizon 2020 Framework Program via the ERC Advanced Grant Origins 83 24 28. G.V., F.E.T. and S.M.E acknowledge support from NSF Astronomy and Astrophysics Research Grant No. 1615897. 

%\bibliography{ref}
%\bibliographystyle{aa}

\end{document}